# Effect of Functionalization on the Properties of Silsesquioxane; a Comparison to Silica

*Marzieh Moradi[a], Bailey M. Woods[a], Hemali Rathnayake[b], Stuart J. Williams[c], Gerold A. Willing[a,]\**

[a] Chemical Engineering, University of Louisville, Louisville, KY 40292, USA
[b] Department of Nanoscience, University of North Carolina at Greensboro, Greensboro, NC 27401, USA
[c] Mechanical Engineering, University of Louisville, Louisville, KY 40292, USA
\* Corresponding author: E-mail address: gerold.willing@louisville.edu

While similar in nature, the properties of silica and silsesquioxane are very different, but little is known about these differences. In this paper, functionalized silsesquioxane microparticles are synthesized by adapting the modified Stöber method and post-functionalized with rhodamine-B. The as synthesized silsesquioxane particles are characterized by a variety of physical and chemical methods. The synthesized particles are amorphous and nonporous in nature and are less dense than silica. While silsesquioxane and silica have some similar physical properties from their siloxane core, the organic functional group of silsesquioxane and the one-half oxygen difference in its structure impact many other properties of these particles like their charging behavior in liquids. These differences not only allow for the ease surface modification as compared to that necessary to modify silica, but also the use in a variety of colloidal systems that due to pH or electrolyte concentrations may not be suitable for silica particles.

## 1. Introduction

Colloidal particles and suspensions have been investigated widely due to their extensive applications from advanced materials to drug delivery [1,2]. Widespread research on colloidal particles has taken place in a number of application areas including coatings [3], assembly of ceramics [4], photonic materials [5], and pharmaceutical materials [6]. Different characteristics of colloidal suspensions and resulting products can be reached by tailoring both the morphology and chemistry of the colloidal particles [7].

The most important properties of a colloidal system such as stability depend on the potential or electric charge of the colloidal particles which is determined by dipolar molecules and adsorption of ions [8]. Interaction forces between colloidal particles can be determined by the potential distribution created by the variation of these molecules and ions between the particles themselves [1,2,9]. Interaction forces are responsible for the stability of particles against coagulation and have an important role in determining properties such as the shelf life, stability, rheology, and the overall behavior in several industrial processes (e.g. mixing and membrane filtration) [9]. Therefore, it is essential to know the charge distribution of a colloidal system to understand the system chemistry and its impact on the interaction between particles.

In colloidal systems, the surface of particles is surrounded by an electric double layer which is also known as the electrical diffuse double layer [8]. This double layer is formed by the particle surface charge and its counter ions, forming an ionic cloud surrounding the particle [9]. The electrical potential of the double layer is commonly referred to as the zeta potential. In fact, zeta potential is the potential difference between the dispersion medium and the stationary layer of fluid associated with the colloidal particle [10].

Silsesquioxanes have gained significant attention due to the fact that they can be easily synthesized and further modified with organic or inorganic functionalities [11-17]. Silsesquioxane refers to structures with a ratio of 1.5 of oxygen to silicon atoms resulting in the general formula of $R-SiO_{1.5}$. In this case, the R can represent a hydrogen or an organofunctional group [18,19]. Based on the chemistry of the silsesquioxanes with an organic-inorganic composition, their properties are combination of ceramic like properties of silica and the soft nature of organic materials [20]. Silsesquioxanes have potential applications as nanoscale fillers in polymer systems for use in adhesives, coatings, composites, and dental fillings [7,21]. They have also been examined for semiconducting devices [22], fuel cells [23], optical devices [24] and sensors [25]. But there has been limited effort aimed at chemical modifications of silsesquioxane particles with reactive organic functional groups.

Recently, Rathnayake et al. [7] made novel benzyl chloride, and benzyl chloride-amine functionalized silsesquioxane (BC-SSQ, BC-A-SSQ) particles by adopting a modified Stöber method. They also post-functionalized the synthesized particles with a Rhodamine-B (RhB) fluorescent tag (BC-SSQ-RhB, BC-A-SSQ-RhB). Silsesquioxane with benzyl chloride functionalization is useful for the tailoring and grafting of a wide variety of materials to the surface including polymer ligands and other precursors. The post-functionalization capabilities of these particles are significantly beneficial for numerous applications in colloid chemistry and nanoscience. Due to the nature of the reactive group functionality which mimics strong dipole-dipole and hydrogen bonding interactions, a colloidal suspension of these particles can assemble into monolayers and hollow colloidasomes on a variety of polymer-coated substrates to create 3D colloidal assemblies [7,26]. Considering the applications of silsesquioxanes in adhesives, coatings and semiconductors, it is very important to determine a wide range of properties for these particles such as their charge, density and porosity along with their morphology.

In this paper, we synthesized the BC-SSQ, BC-A-SSQ, BC-SSQ-RhB and BC-A-SSQ-RhB particles based on the Rathnayake et al. method [7] and report our efforts to fully characterize these microparticles. Zeta potential measurements of the particles were performed over a range of pH to determine the charging behavior of the particles and to find their isoelectric point. The morphology and surface analysis of the particles was examined by a helium ion microscope (HIM) and a scanning electron microscope (SEM). Average size of the particles was measured using a particle size analyzer. Density was measured by a helium gas pycnometer and specific surface area of particles was measured using a BET instrument.

## 2. Methods and Materials

### 2.1 Materials

Para-(chloromethyl)phenyltrimethoxy silane was purchased from Gelest Inc. Potassium carbonate (ACS reagent) was purchased from Fischer Scientific. 3-aminopropyltriethoxy silane (3-APT), anhydrous ethanol (200 proof), ammonium hydroxide (28%), and Rhodamine-B carboxylic acid were purchased from VWR international.

### 2.2. Particle Synthesis

For synthesizing BC-SSQ particles, 10 mL of anhydrous ethanol and 4 mL of 28% ammonium hydroxide were mixed on a magnetic stirrer for 5 minutes. Then 1.8 mL of benzyl chloride trimethoxy silane was added to the reaction at a rate of 0.08 mL/minute and was allowed to stir for 18 hours. Particles were separated by centrifuging at 3000 rpm for 20 minutes. Then, particles were washed multiple times with ethanol followed by distilled water to remove any impurities. Finally, particles were dried under the hood for 48 hours to yield BC-SSQ particles.

For synthesizing BC-A-SSQ particles, 65 mL of anhydrous ethanol and 3.5 mL of 28% ammonium hydroxide were mixed on a magnetic stirrer for 5 minutes. Then 2 mL of 3-APT silane was added to the flask dropwise followed immediately by 1 mL of benzyl chloride trimethoxy silane at a rate of 0.08 mL/minute and was allowed to stir for 18 hours. Then the BC-A-SSQ particles were washed and dried the same way as BC-SSQ particles.

In order to functionalize the as synthesized particles with Rhodamine-B, 500 mg of particles were dispersed in 30 mL of anhydrous ethanol and were mixed on a magnetic stirrer until the particles were dispersed completely. Then 79 mg of potassium carbonate was added followed by 230 mg of Rhodamine-B and was allowed to stir for 18 hours. In order to keep the reaction away from visible light, the reaction flask was covered with aluminum foil. The same procedure was used to wash and dry the particles.

### 2.3. Structure, Morphology, and Size

In this work, the morphology of particles was investigated using a Zeiss Auriga Crossbeam FIB-FESEM scanning electron microscope and a Zeiss Orion helium ion microscope. A TESCAN scanning electron microscope (SEM) equipped with energy-dispersive X-ray spectroscopy (EDX) was used for surface analysis of the particles. X-ray diffraction (XRD) measurements were carried out on a Bruker D8 Discover diffractometer. Patterns were recorded over the range from 10 to 90° ($2\theta$) in steps of 0.02° with a scan speed of 2s at each step. A Brookhaven 90 Plus-Zeta particle size analyzer was used to measure the average sizes of particles and their size distributions.

### 2.4. Zeta Potential

The Brookhaven 90 Plus-Zeta particle size analyzer was used to measure the zeta potential of colloidal particles as well. Zeta potential of particles was measured at 0.01 volume percent concentration of particles in solutions with pH ranging from 3 to 11. Acidic and basic solutions were made by adding proper amount of nitric acid or potassium hydroxide to DI water. Each sample was sonicated and well-suspended in the pH solution before zeta potential measurements. These measurements also revealed the isoelectric point of the particles. A Nikon Eclipse Ti inverted microscope was then used to observe any agglomerates that formed during settling.

### 2.5. Density

Particle density is required for sedimentation analysis, or calculations involving volumes or mass of particles. The particle density of different SSQ particles was measured by a AccuPyc II 1340 Pycnometer which is a fully automatic gas displacement pycnometer. Particles were first dried in the desiccator to obtain true sample mass and to avoid the distorting effect of water vapor on the volume measurement. Mass was determined by precisely weighing particles on an analytical balance. Knowing the mass of the particles, the density was measured by the pycnometer using a 1.0 cm$^3$ cup.

## 2.6. Specific surface area

After drying the particles in a desiccator, particles were degassed using a SmartPrep degasser to remove any gas trapped in the pores and on the other surfaces of the particles. To degas, particles were heated to 120°C and degassed for 2 hours. After degassing, a TriStar 3000 gas adsorption analyzer was used to measure the surface area and porosity of the particles by measuring gas ($N_2$) adsorption.

## 3. Results
### 3.1. Morphology

In the literature, silsesquioxanes have been reported to have a variety of different structures, including a random structure, ladder structure, cage structures, and partial cage structure [13,27].

The X-ray diffraction patterns of the as synthesized BC-SSQ and BC-A-SSQ microparticles are shown in Fig. 1. The XRD spectral traces show that a broad peak is present at $2\theta = 10–30°$ for both particles, verifying them to be of amorphous structure [28,29].

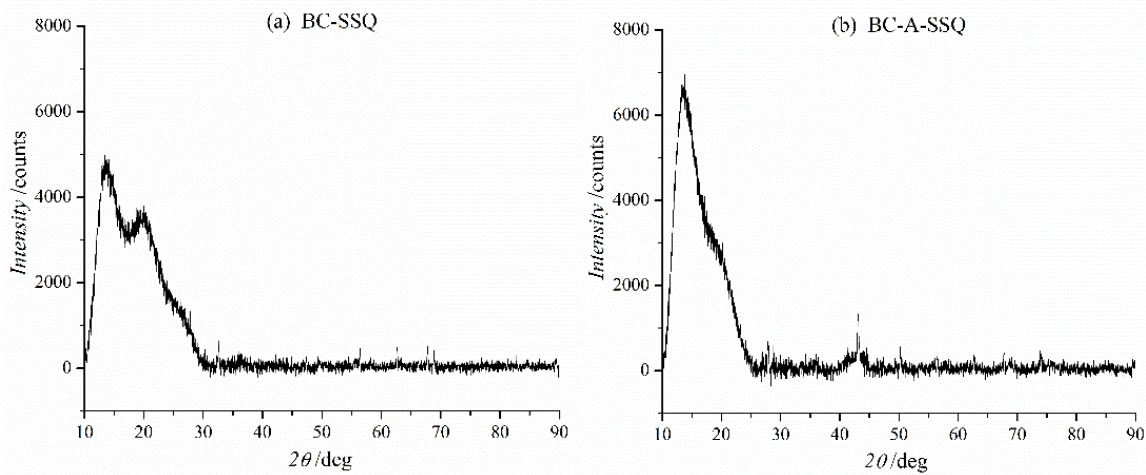

**Fig. 1.** XRD patterns for amorphous (a) BC-SSQ and (b) BC-A-SSQ

SEM and HIM images of BC-SSQ and BC-A-SSQ particles are shown in Fig. 2 and Fig. 3. These images are taken after 4 hours of reaction. As it is clearly seen in these pictures, both particles are mostly spherical with a smooth surface which is in agreement with N. Neerudu's results [7].

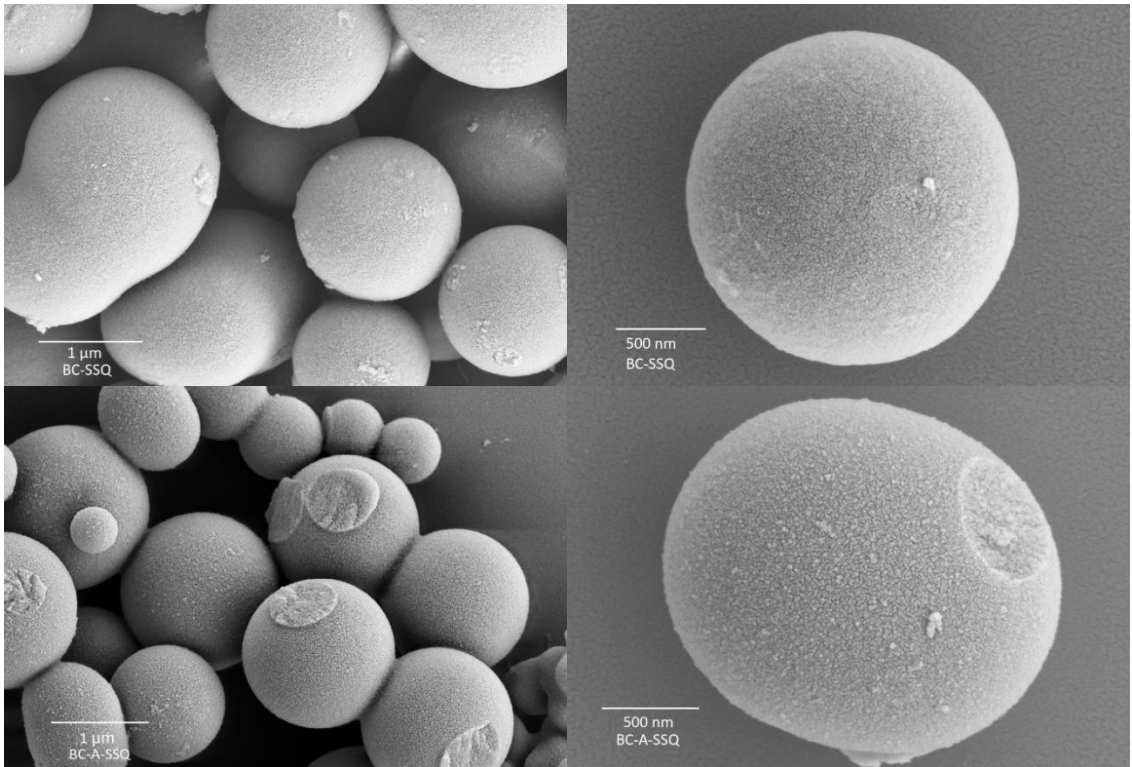
**Fig. 2.** SEM images of BC-SSQ and BC-A-SSQ

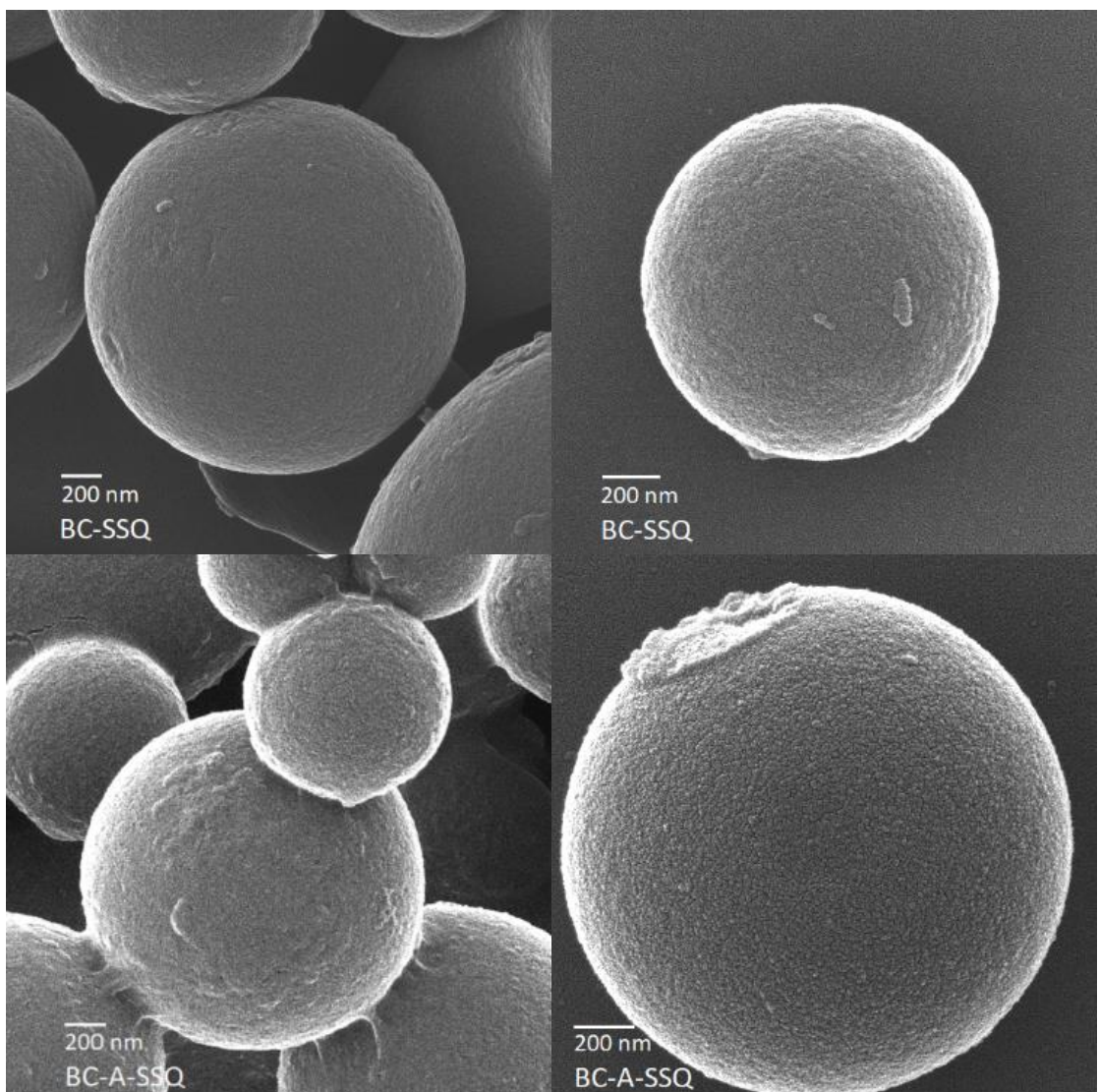
**Fig. 3.** HIM images of BC-SSQ and BC-A-SSQ

In Table 1, the EDX descriptive statistics of the average amount of different elements in the surface layer of BC-SSQ and BC-A-SSQ particles are presented. One may notice that the chlorine amount in the surface layer is much higher in BC-SSQ particles compared to the BC-A-SSQ particles. This is expected as the amine functionalization will cover and replace many of the exposed benzyl chloride groups on the particle surface.

**Table 1.** Descriptive statistics of the amount of different elements in the surface layer of BC-SSQ and BC-A-SSQ particles.

| Element | Atomic % | | Ratios to Si | |
|---|---|---|---|---|
| | BC-SSQ | BC-A-SSQ | BC-SSQ | BC-A-SSQ |
| Si | 27.77 | 26.94 | 1 | 1 |
| O | 40.52 | 43.33 | 1.46 | 1.61 |
| Cl | 31.71 | 17.83 | 1.14 | 0.66 |
| N | 0 | 11.90 | 0 | 0.44 |

Particle size measurements by DLS method, as shown by volume in Fig. 4, showed BC-SSQ and BC-A-SSQ particles with sizes in the range of 600-2700 nm and 1400-3600 nm, respectively. The median size was determined for BC-SSQ particles as 1200 nm and for BC-A-SSQ as 2300 nm.

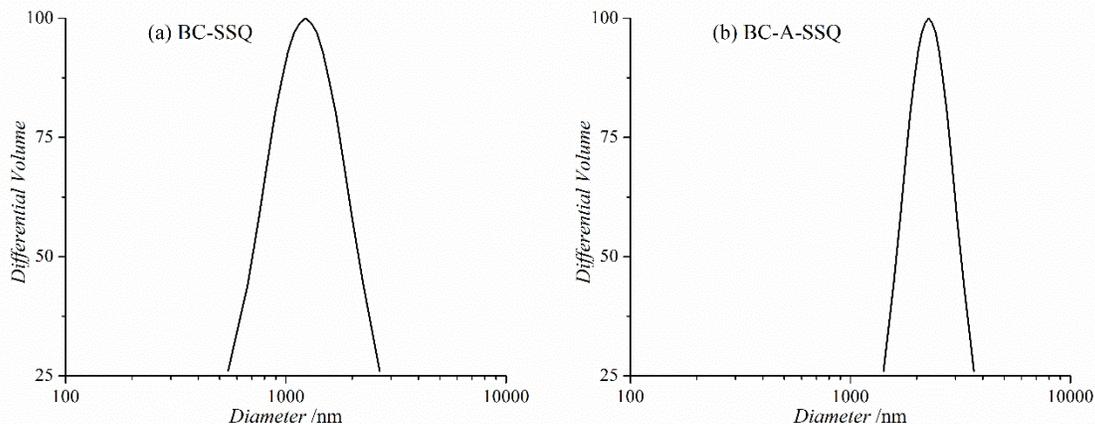

**Fig. 4.** Lognormal differential distributions of particle sizes weighted by volume for (a) BC-SSC and (b) BC-A-SSQ particles

### 3.2. Zeta Potential

Fig. 5 shows zeta potential measurements as a function of pH for the BC-SSQ and BC-SSQ particles. It is noted that both particles are positively charged at lower acidic pH while bifunctional BC-A-SSQ particles have a higher charge at a lower pH. This confirms the presence of amino groups in the BC-A-SSQ particles surface in their protonated form. At higher pH, that is towards the basic end of the scale, the value of the zeta potential changes from positive to negative. The isoelectric point of the BC-SSQ and BC-A-SSQ were found to be near pH 4 and 7, respectively, while the isoelectric point of silica has been reported to be between pH 1.7 and pH 2.5 [30,31]. This makes SSQ favorable for conditions where highly acidic solutions cannot be used due to materials or safety concerns. It should also be noted that RhB functionalization did not significantly affect the charge of particles [32]. This is extremely valuable for imaging in confocal microscopy because one can easily fluorescently label these SSQ particles and study the structure of agglomerates without changing their isoelectric point.

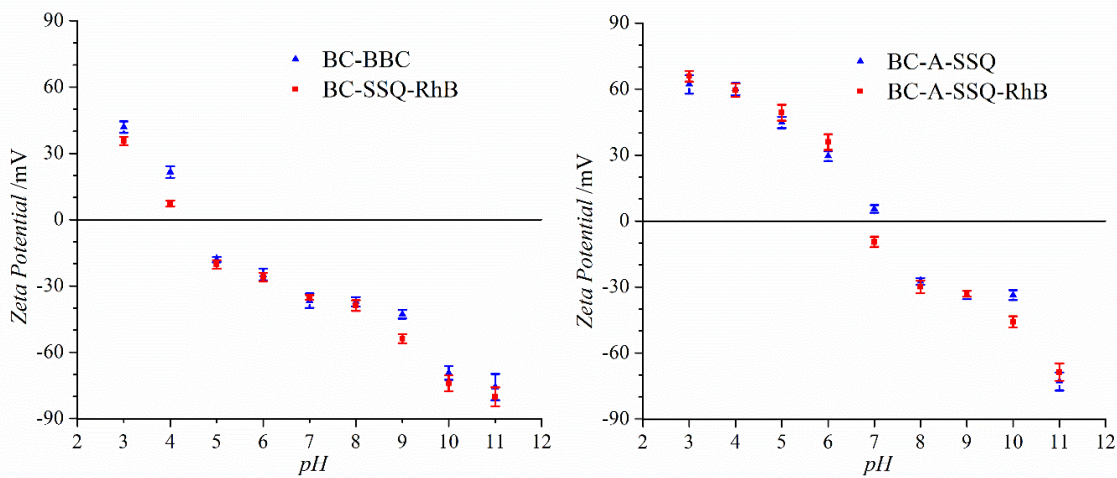

**Fig. 5.** Plot of effective zeta potential (ξ) of the BC-SSQ, BC-SSQ-RhB and BC-A-SSQ, BC-A-SSQ-RhB microparticles as a function of pH

Fig. 6 shows microscope images of the BC-SSQ samples at pH 3, 4, and the isoelectric point. As it can be seen in this figure, by decreasing the pH from isoelectric point to 3 which increases the electric charge of the particles, we observe that the particles are more dispersed during settling. These images were analyzed using ImagaJ. For this purpose, images were converted into 8-bit, and then an Iso-Data algorithm threshold was set to differentiate the particles from the backdrop [33]. A binary close operation and a watershed separation was performed on the thresholded images. The watershed function visually separated semi-agglomerated particles, and ImageJ output the particle count, total blob area and average blob size which are listed in Table 2. As the pH increases, the number of aggregates, total blob areda and average blob size increase accordingly. This is consistent with the lower stability and higher tendency of the particles to agglomerate at their isoelectric point. Such behavior for silica particles would not be observed unless the pH was less than 2. It should be noted that this is a result of the BC-SSQ particles taking on a strong positive electric charge below pH values of 4, unlike silica particles that would not take on a positive charge unless the pH value is less than 2.

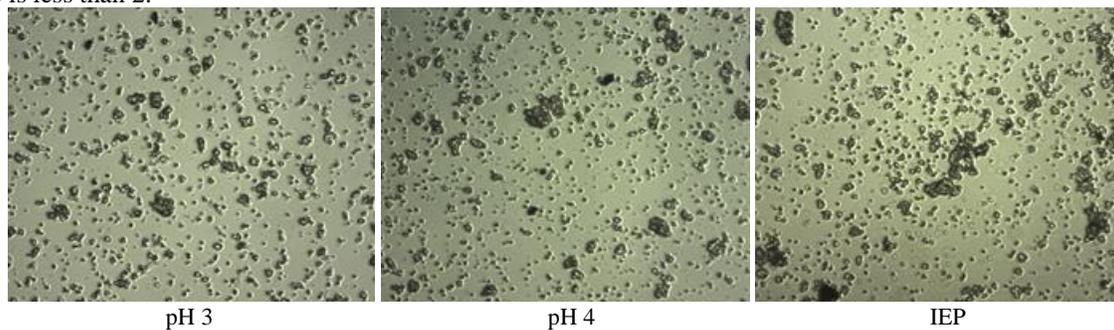

pH 3          pH 4          IEP

**Fig. 6.** Images of BC-SSQ particles at pH 3, 4 and isoelectric point after 4 hrs using inverted microscopy (20x).

**Table 2.** ImageJ analysis data of BC-SSQ samples

| pH  | Count | Total blob area | Average blob size |
|-----|-------|-----------------|-------------------|
| 2   | 337   | 4921            | 13.635            |
| 3   | 347   | 6063            | 17.473            |
| IEP | 394   | 8967            | 22.759            |

### 3.3. Density
The measured particle density for BC-SSQ and BC-A-SSQ was 1.68±0.05 and 1.48±0.04 g/cm$^3$ at 22°C, respectively. This is lower than the density of silica which is reported to be 2.196 g/cm$^3$ [34]. This result is not unexpected as the structure of the silsesquioxane is highly amorphous and has a random cage structure which is more open than the structure of amorphous silica.

### 3.4. Specific surface area
The BET surface area ($S_{BET}$) of BC-SSQ and BC-A-SSQ particles with size of 1.2 µm and 2.3 µm was 3.83±0.07 m$^2$/g and 4.72±0.04 m$^2$/g, respectively. The average pore width of BC-SSQ and BC-A-SSQ particles was 7.97 nm and 8.76 nm, respectively. Fig. 7 shows the physisorption and pore size distribution of the particles. It can be seen that both BC-SSQ and BC-A-SSQ particles exhibit a typical type II isotherm [35]. Considering the size of the pores and type II isotherms, both types of particles possess a nonporous structure [35,36]. The gradual curvature and less distinctive Point B indicate a significant amount of overlap of monolayer coverage and the onset of multilayer adsorption. The thickness of the adsorbed multilayer generally appears to increase without limit when $P/P_0=1$.

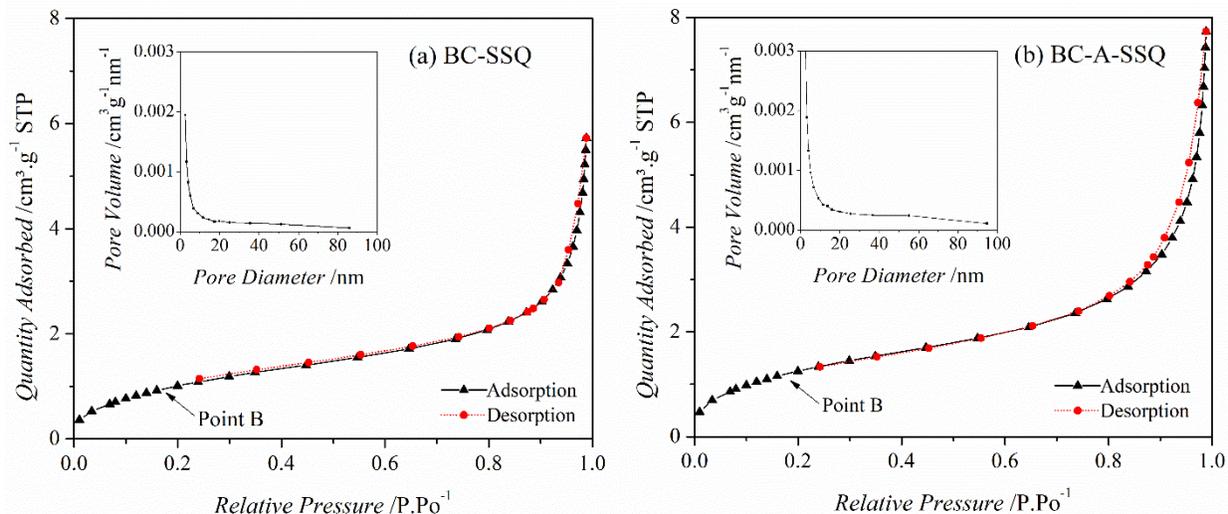
**Fig. 7.** N$_2$ adsorption-desorption isotherm and pore size distribution of (a) BC-SSQ and (b) BC-A-SSQ

### 3.5. General Overall Structure
Based on the characterization results, all four types of functionalized silsesquioxane particles tested in this study were spherical and had a nonporous, amorphous structure. It was also shown that the silsesquioxane particles were less dense than silica. Compared to silica, the potential colloidal stability of silsesquioxane was different due to the difference in the structure and surface charges. Benzyl chloride and amine functionalization of different silsesquioxane particles significantly affected the charge of particles. The isoelectric points of all four types of silsesquioxane particles were significantly higher than the isoelectric point of silica. These differences may make these particles more viable in applications where an isoelectric point closer to a neutral pH or a lower density particle is necessary.

Silsesquioxane is most often treated the same as silica. But in this study, it was shown that although silsesquioxane and silica have a number of similarities in their physical properties such as thermal stability and rigidity which result from their siloxane core, the one-half oxygen difference in the silsesquioxane structure along with its' organic functional groups significantly impact many other properties of these particles. The more open structure of silsesquioxane makes it less dense when compared to silica. The difference in the number of oxygen molecules, as well as the type of the functional groups in the structure of the silsesquioxane particles as compared to silica particles, changes the overall charge of the particles in the solution and results in a significantly higher isoelectric point. Silsesquioxane particles can be positively charged in higher range of pH while silica is almost always negatively charged due to its low isoelectric point. Charging behavior of colloids impacts the stability of the suspension and is very important for designing a particular colloidal suspension based on the desired properties of that suspension. The post-functionalization capabilities of silsesquioxanes, based on both silane and halogen chemistry, are also beneficial for numerous applications in colloid science and ultimately give them a wider range of application when compared to silica.

### 4. Conclusion
Silsesquioxane microparticles with different functional groups were synthesized by the modified Stöber method and post-functionalized with rhodamine-B. Both BC-SSQ and BC-A-SSQ particles were spherical with a smooth surface. The XRD results showed that both particles had amorphous structure. The isoelectric point of the BC-SSQ particles and BC-A-SSQ were near pH 4 and 7. Density of BC-SSQ and BC-A-SSQ was 1.68±0.05 and 1.48±0.04 g/cm$^3$ at 22ºC, respectively. Both BC-SSQ and BC-A-SSQ were nonporous with an average pore width of 7.97 nm and 8.76 nm and BET surface area of 3.83±0.07 m$^2$/g and 4.72±0.04 m$^2$/g, respectively. While these particles share some similarities with silica particles, the elimination of the one-half oxygen atom and the inclusion of the organic ligands into the overall structure changes the physical and chemical properties in significant ways. These property changes will not only allow silsesquioxane particles to be used in a wider variety of applications and but will also provide a greater means for tailoring system properties for desired behaviors and products.

**Funding information**
This work was financially supported by a grant from NASA EPSCoR (Grant No. NNX14AN28A).

**Data Availability**
The raw data required to reproduce these findings are available and can be requested from the authors. The processed data required to reproduce these findings are available and can be requested from the authors.
**Conflicts of interest** The authors declare that they have no conflict of interest.